\documentstyle[psfig]{elsart}

\def\FIG{}
\def\be{\begin{equation}}
\def\ee{\end{equation}}
\def\P{\partial}
\def\q{\\}

\begin{document}
\pssilent
\begin{frontmatter}
\title{Collective behavior of interacting self-propelled particles}

\author{Andr\'as Czir\'ok and Tam\'as Vicsek}

\address{Dept. of Biological Physics, Eotvos University, Budapest, Hungary}

\begin{abstract}
We discuss biologically inspired, inherently non-equilibrium self-propelled
particle models, in which the particles interact with their neighbours
by choosing at each time step the local average direction of motion.
We summarize some of the results of large scale simulations and theoretical
approaches to the problem.
\end{abstract}

\end{frontmatter}

\section{Introduction}

\def\v{\vec{v}}
\def\x{\vec{x}}
\def\u{\vec{u}}
\def\e{\vec{e}}
\def\ro{\rho}
\def\be{\begin{equation}}
\def\ee{\end{equation}}
\def\P{\partial}
\def\d{\hbox{d}}

The collective motion of various organisms like the flocking of birds,
swimming of schools of fish, motion of
herds of quadrupeds (see, e.g. \cite{Par82,PE99} and references therein) 
migrating bacteria \cite{AH91}, molds
\cite{RNSLN99}, ants \cite{RMC95} or pedestrians \cite{HKM97} is a fascinating
phenomenon of nature.  We address the question whether there are some global,
perhaps universal features of this type of behavior when many organisms are
involved and parameters like the level of {\em perturbations} or the mean {\em
distance} between the individuals are changed.

These studies are also motivated by recent developments in areas related to
statistical physics.  Concepts originated from the physics of phase transitions
in equilibrium systems \cite{Ma85} such as scale
invariance and renormalization have also been shown to be useful in the
understanding of various non-equilibrium systems, typical in our natural and
social environment.  Motion and related transport phenomena represent further
characteristic aspects of many non-equilibrium processes and they are 
essential features of most living systems.

To study the collective motion of large groups of organisms, the concept of
{\em self propelled particle} (SPP) models was introduced in \cite{VCBCS95}. As
the motion of flocking organisms is usually controlled by interactions with
their neighbors \cite{Par82}, the SPP models consist of locally interacting
particles with an intrinsic driving force, hence with a finite steady velocity.
Because of their simplicity, such models represent a statistical approach
complementing other studies which take into account more details of the actual
behavior \cite{HKM97,Rey87,SSMHS96}, but treat only a moderate number of
organisms and concentrate less on the large scale behavior.

In spite of the analogies with {\em ferromagnetic} models, the general behavior
of SPP systems can be quite different from those observed in equilibrium
magnets.  In particular, equilibrium ferromagnets possessing continuous
rotational symmetry do not have ordered phase at finite temperatures in two
dimensions \cite{MW66}.  However, in 2d SPP models an ordered phase can
exist at finite noise levels (temperatures) as it was first demonstrated by
simulations \cite{VCBCS95,CSV97} and explained by a theory of flocking
developed by Toner and Tu \cite{TT95,TT98}.  Further studies revealed that
modeling collective motion leads to interesting specific results in all of the
relevant dimensions (from 1 to 3) \cite{CBV99,CVV99}.

\section{Discrete models}\label{sec:2}
	\paragraph*{The 2d system.}


The simplest model, introduced in \cite{VCBCS95}, consists of particles moving
on a plane with periodic boundary condition. The particles are characterized by
their (off-lattice) location $\vec{x}_i$ and velocity $\vec{v}_i$ pointing in
direction $\vartheta_i$. The self-propelled nature of the particles is
manifested by keeping the magnitude of the velocity fixed to $v_0$. 
Particles interact through the following local rule:
at each time step a given particle
assumes the {\em average direction of motion} of the particles in its local
neighborhood $S(i)$ (e.g., in a circle of some given radius centered at the position of the $i$th particle) with some uncertainty, as described by
\begin{equation}
	\vartheta_i (t+\Delta t) = \langle \vartheta(t) \rangle_{S(i)} + \xi,
\label{SPP2d_EOM}
\end{equation}
where the noise $\xi$ is a random variable with a uniform distribution in the
interval $[-\eta/2,\eta/2]$.  The locations of the particles are updated as
\begin{equation}
	\vec{x}_i(t+\Delta t) = \vec{x}_i(t) + \vec{v}_i(t)\Delta t
\label{SPP2d_update}
\end{equation}
with $\vert \vec{v}_i\vert=v_0=const$.

The model defined by Eqs. (\ref{SPP2d_EOM}) and (\ref{SPP2d_update}) is a
transport related, non-equilibrium analog of { ferromagnetic} models
\cite{Sti83}.  The analogy is as follows: the Hamiltonian tending to align the
spins in the same direction in the case of equilibrium ferromagnets is replaced
by the rule of aligning the direction of motion of particles, and  the
amplitude of the random perturbations can be considered proportional to the
temperature for $\eta\ll 1$.  From a hydrodynamical point of view, in SPP
systems the momentum of the particles is {\em not} conserved and -- as was
pointed out in \cite{TTU98} -- the Galilean invariance is broken. Thus, the flow
fields emerging in these models can considerably differ from the usual behavior
of fluids.

	\paragraph*{Critical exponents.}

\begin{figure}
\FIG{\psfig{figure=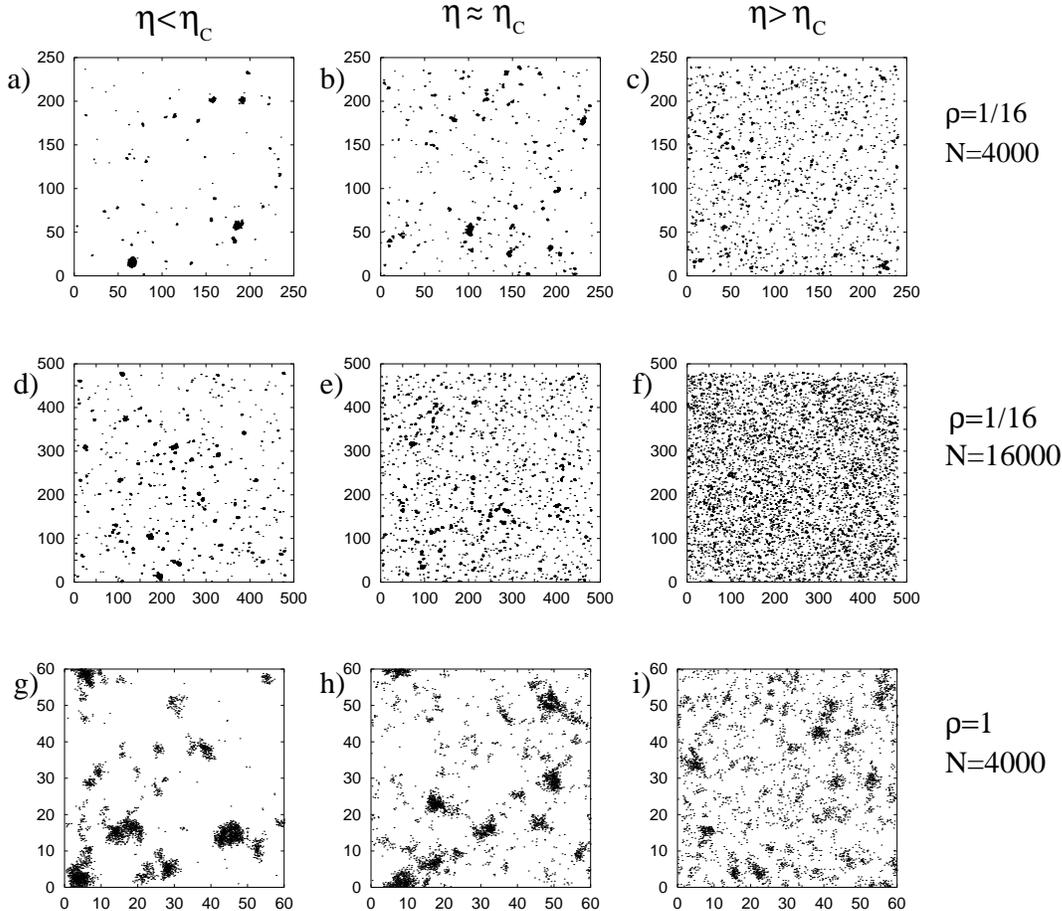,width=14cm,angle=-90}}
\caption{\sl
	Typical configurations of the positions of SPPs displayed for various
	values of density and noise. For small enough noise coherently moving
	clusters can be identified in the system, which gradually disappear
	as the noise amplitude is increased. The noise amplitudes applied
	in the simulations were the following: $\eta=0.2$ (a,d); $\eta=0.4$ 
	(b,e); $\eta=0.8$ (c,f); $\eta=0.6$ (g); $\eta=1.2$ (h) and $\eta=1.8$
	(i). The corresponding order parameter values are: $\phi=0.75$ (a,d);
	$\phi=0.19$ (b); $\phi=0.025$ (c); $\phi=0.3$ (e,h); $\phi=0.05$ (f,i)
	and $\phi=0.85$ (g). The scales are given in units of the interaction
	length.
	}
\label{sppfig2}
\end{figure}

The model defined through Eqs.  (\ref{SPP2d_EOM}) and (\ref{SPP2d_update}) 
with circular interaction range was
studied by performing large-scale simulations.
Due to the simplicity of the model, only two control parameters should be
distinguished: the (average) density of particles $\varrho=N/L^2$ (where
$N$ is the number of particles and $L$ is the system size in units of the
interaction range) and the amplitude
of the noise $\eta$. Depending on the value of these parameters the model
can exhibit various type of behaviors as Fig. \ref{sppfig2} demonstrates.

\begin{figure}
\centerline{
\hbox{
	\FIG{\psfig{figure=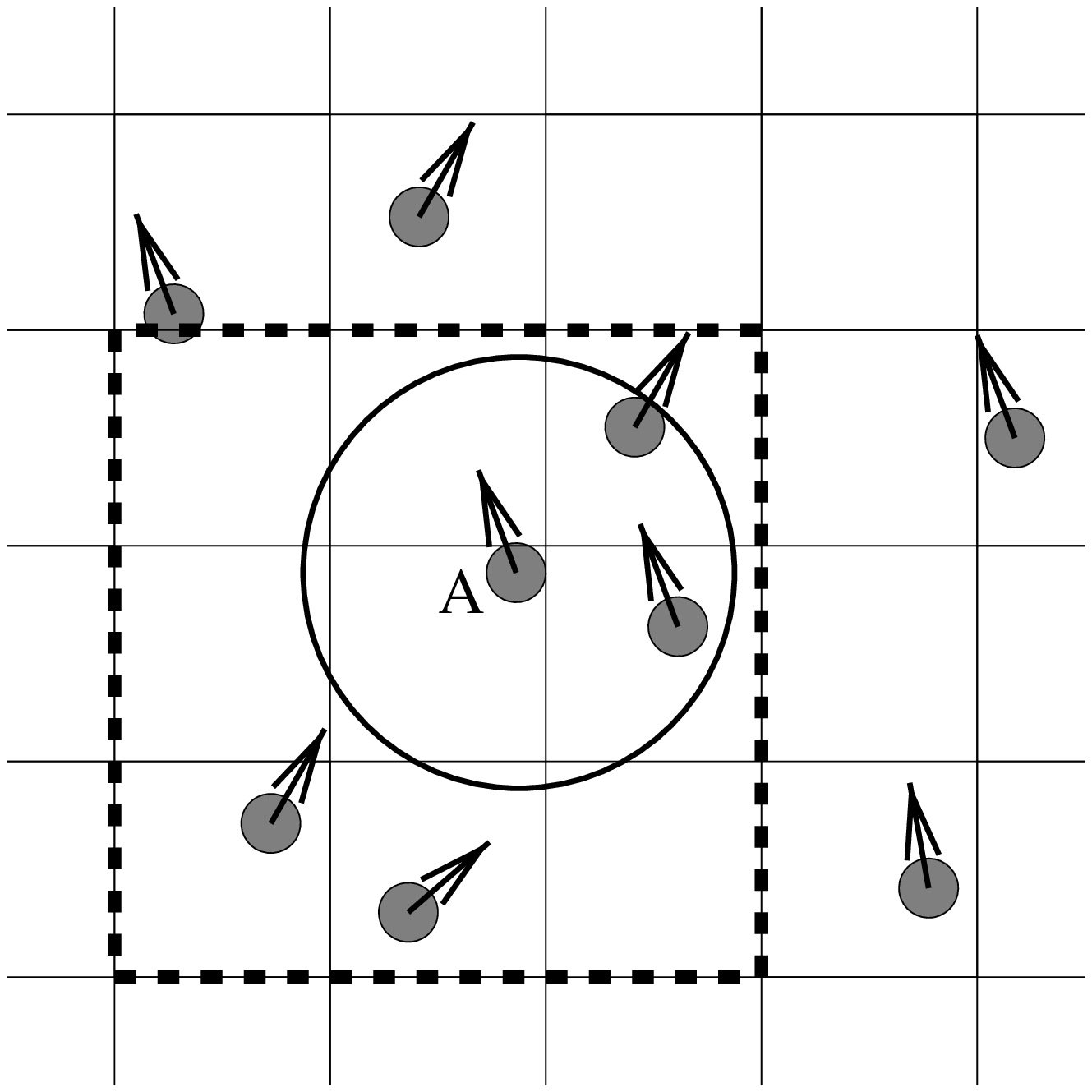,height=6.0truecm,angle=-90}}
	\hskip 0.5cm
	\FIG{\psfig{figure=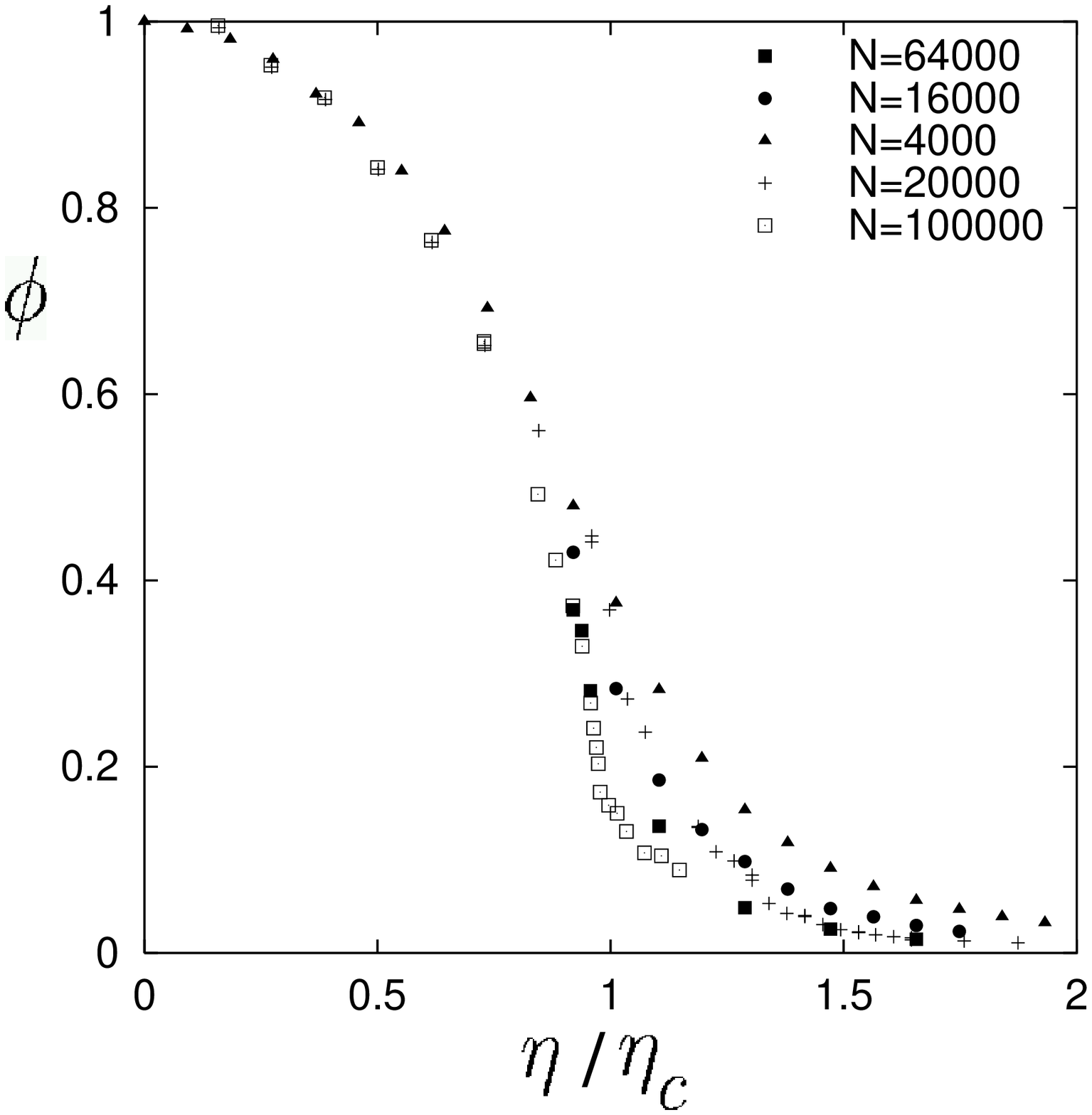,height=6.0truecm,angle=-90}}
}}
\caption{\sl
	Left: Schematic illustration of the various realizations of the SPP
	model. Particles move off-lattice on a plane and interact with other
	particles located in the surrounding,  which can be either a
	circle ($S_1$) or 9 neighboring cells in an underlying lattice ($S_2$).
	We plot these interaction areas for particle $A$ with a solid and
	dashed line, respectively.  
	Right: The average momentum of the system in the steady state vs the
	rescaled noise amplitude $\eta/\eta_c$ for various interaction
	ranges, densities and system sizes. Open symbols refer to interaction 
	type $S_1$ and $\varrho = 1$ while filled symbols refer to interaction 
	type $S_2$ and $\varrho=2$. 
	The order present at small $\eta$ disappears in a continuous manner
	reminiscent of second order phase transitions for $N\rightarrow\infty$.
	}
\label{SPP_univ}
\end{figure}

For the statistical characterization of the system a well-suited order
parameter is the magnitude of the {\em average momentum} of the particles:
$ \phi\equiv{1\over N} \left\vert \sum_j \vec{v}_j \right\vert. $
This measure of the net
flow is non-zero in the ordered phase, and vanishes (for an infinite system) in
the disordered phase.  Since the simulations were started from a random, disordered
configuration, $\phi(t=0)\approx 0$.  After some relaxation time a steady state
emerges indicated by the convergence of the cumulative average $(1/\tau)
\int^\tau_0 \phi(t)dt$.  The stationary values of $\phi$ are plotted in
Fig.~\ref{SPP_univ} vs $\eta$ for various parameters.
For weak noise the model displays an ordered motion, i.e. $\phi\approx 1$, 
which disappears in a continuous manner by increasing $\eta$.
As $L \rightarrow \infty$, the numerical results show the
presence of a phase transition described by
\begin{equation}
\phi(\eta)\sim \cases{
         \Bigl({\eta_c(\varrho) - \eta\over \eta_c(\varrho)}\Bigr)^\beta
                & for $\eta<\eta_c(\varrho)$ \cr
        0  & for $\eta>\eta_c(\varrho)$ \cr
    },
\label{scale}
\end{equation}
where $\eta_c(\varrho)$ is the critical noise amplitude that separates
the ordered and disordered phases and $\beta_{2d}=0.42\pm0.03$,
was found \cite{CSV97} to be
{\em different} from the the mean-field value $1/2$. This large-scale 
behavior does not depend on the specific choice of the microscopic details
as Fig.~\ref{SPP_univ} demonstrates.

As a further analogy with equilibrium phase transitions, the fluctuations of
the order parameter also increase on approaching the critical line 
\cite{CSV97}.  The tails of the curves are symmetric, and decay as
power-laws 
with an exponent $\gamma$ close to $2$, which value is, again, different
from the mean-field result. 

Next we discuss the role of density. 
The
long-range ordered phase is present for any $\varrho$, but for a fixed
value of $\eta$, $\phi$ vanishes with decreasing $\varrho$.
The critical line $\eta_c(\varrho)$ in the $\eta-\varrho$ 
parameter space was found to follow
\begin{equation}
\eta_c(\varrho)\sim\varrho^\kappa,
\label{kappa}
\end{equation}
with $\kappa = 0.45 \pm 0.05$ for $\varrho\ll1$.
This critical line  is qualitatively different from that of
the diluted ferromagnets, since here
the critical density at
$\eta\rightarrow0$ (corresponding to the percolation threshold for
diluted ferromagnets, see, e.g., \cite{Sti83}) is vanishing.

\begin{figure}
	\FIG{\centerline{ \psfig{figure=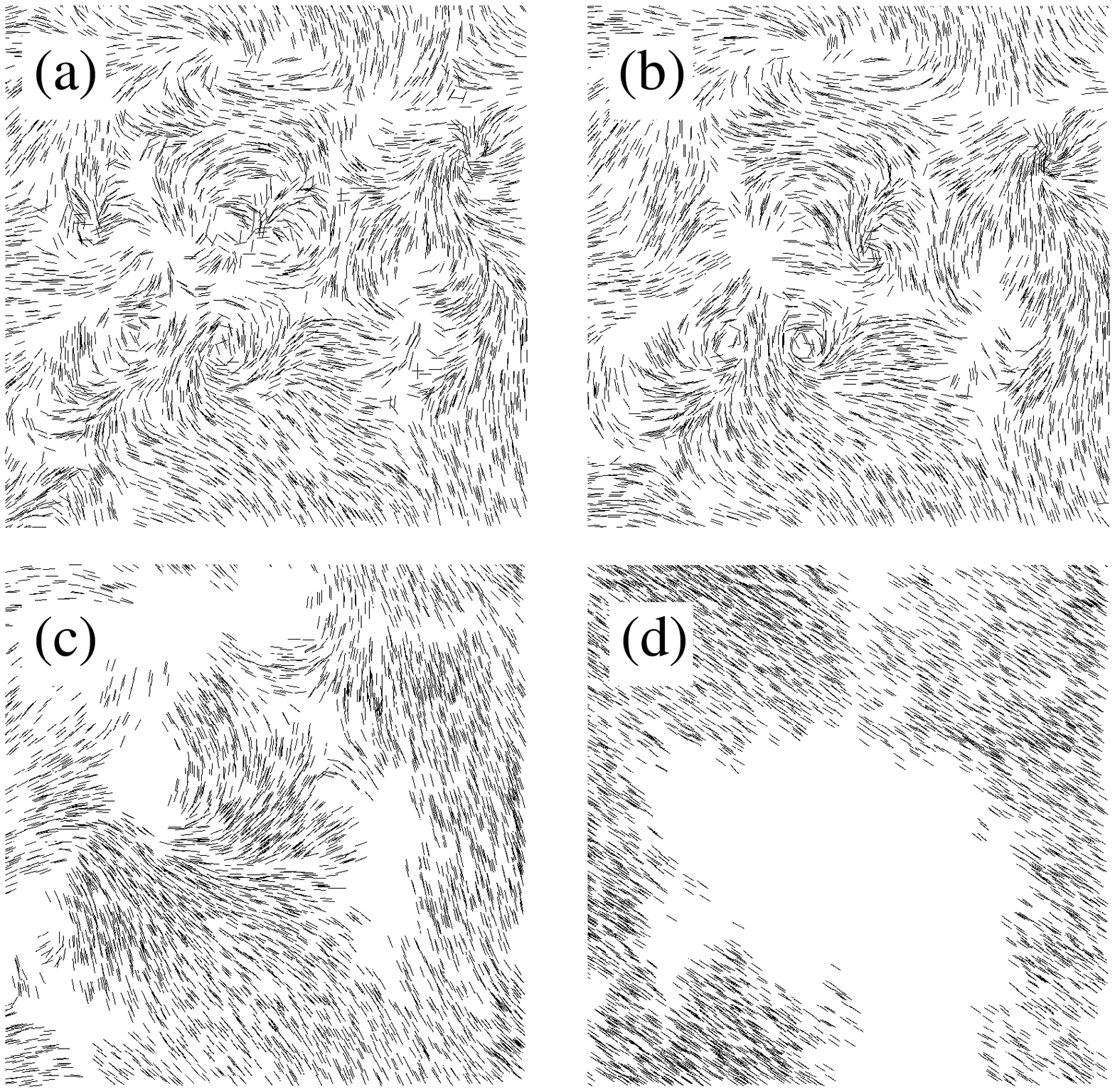,height=12truecm} }}
\caption{\sl
	Snapshots of the time development of a system with  $N=4000$, $L=40$
	and $v_0=0.01$ at $50$ (a), $100$ (b), $400$ (c) and $3000$ (d)
	steps.  First the behavior is reminiscent of the
	equilibrium $XY$ model, where the long range order is missing since
	vortices are present in the system.  Here the vortices are unstable,
	and finally a self-organized long-range order develops.
	(After \cite{CSV97}.)
	}
\label{SPP_5}
\end{figure}

These findings indicate that SPP systems can be quite well characterized
using the framework of classical critical phenomena, but also show surprising
features when compared to the analogous equilibrium systems.  The velocity
$v_0$ provides a control parameter which switches between the SPP behavior
($v_0>0$) and an $XY$ ferromagnet ($v_0=0$).  Indeed, for $v_0=0$
Kosterlitz-Thouless vortices \cite{KT73} can be observed in the system, which
are {\em unstable} (Fig.~\ref{SPP_5}.) for any nonzero $v_0$
investigated in \cite{CSV97}.

	\paragraph*{Correlation functions.}

\def\vr{\vec{r}}

Beside the calculation of the order parameter, it is insightful to characterize
the configurations with correlation functions, such as the velocity-velocity
correlation function
\be
C(\vr)=\left\langle \vec{v}(\vr+\vr',t)\vec{v}(\vr',t) \right\rangle_{\vr',t}
\ee
where $\vec{v}(\vr,t)$ is the coarse-grained velocity field and the average
is taken over all possible values of $\vr'$ and $t$. The system is ordered
on the macroscopic scale if 
$C_\infty=\lim_{\vert \vr \vert \rightarrow \infty}C(\vr)>0$. The decay of the
fluctuations is given by the connected piece of the correlation function
$C_C(\vr)$, defined as $C_C(\vr)=C(\vr)-C_\infty$.

One of the major result of the analysis of Toner and Tu \cite{TT98} was that 
the attenuation of fluctuations is {\em anisotropic} in a SPP flock. In 
particular, they predicted that within a certain length scale 
$\vert\vr\vert<\ell$, $C(\vr)$ decays as 
\be
C(\vr)\sim r_\perp^{-2/5}
\label{2d_corr_pred}
\ee
with $r_\perp$ and $r_{||}$ being the orthogonal and parallel projection of 
$\vr$ relative to the average direction of motion $\left\langle \vec{v}(\vr',t)
\right\rangle_{\vr'}$, respectively.

\begin{figure}
\centerline{\psfig{figure=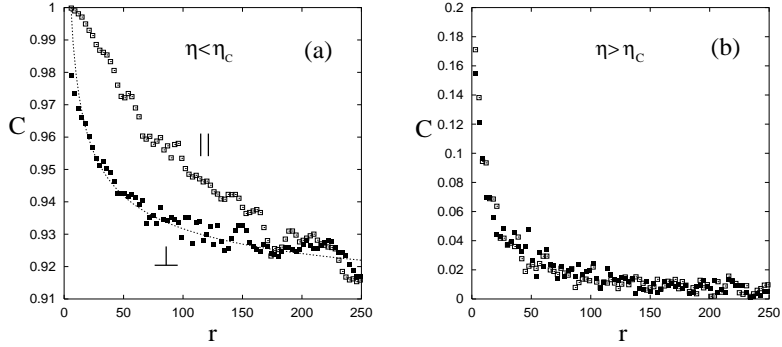,width=12cm,angle=-90}}
\caption{\sl
  Equal time velocity-velocity correlation functions characterizing the
  decay of correlations parallel and perpendicular to the average direction
  of motion of the flock. $C_{||}$ and $C_\perp$ were calculated for systems with
  circular interaction range, $N=4000$, $L=480$ and $\eta=0.02$ (a), 
  $\eta=0.32$ (b). At these parameter values the critical noise amplitude is
  at $\eta_c\approx0.2$. The solid line in (a) represents a power-law fit
  on $C_C$ with the predicted exponent $-2/5$.
}
\label{fig_2d_corr}
\end{figure}

We calculated the equal-time velocity-velocity correlation function in
$(r_\perp$, $r_{||})$ base for various time moments and averaged them to obtain
$C(\vr)$. To demonstrate the predicted
anisotropy of the correlation functions, in Fig.~\ref{fig_2d_corr} curves are
plotted which represent averages of points for which either the 
$r_\perp > r_{||}$ or the $r_\perp < r_{||}$ relation holds:
\begin{eqnarray}
C_\perp(r)=\left\langle C(\vr) \right\rangle_{\vert\vr\vert=r, r_\perp>r_{||}},\hskip5mm
C_{||}(r)=\left\langle C(\vr) \right\rangle_{\vert\vr\vert=r, r_\perp<r_{||}}.
\end{eqnarray}
For $\eta<\eta_c$ the anisotropy of the correlation functions is clearly seen,
while the behavior of $C_\perp$ is consistent with the predicted 
(\ref{2d_corr_pred}).
In contrast, for $\eta>\eta_c$ the curves vanish for large $r$ and the system
is isotropic as expected.

	\paragraph*{Role of boundary conditions.}

Simulations \cite{DHT95,Hem95,CBCV96}
with {\it reflective} boundaries and a short range repulsive force 
(constraining the maximal local density in the system) pointed to
the importance of the boundary conditions in SPP models.

\begin{figure}\label{spp_mosogep}
\centerline{\FIG{\psfig{figure=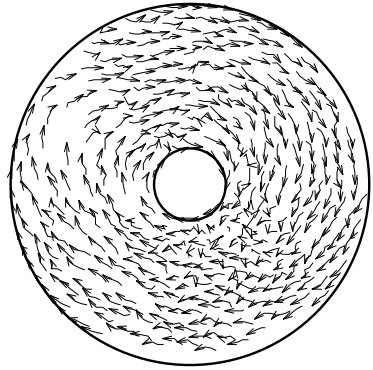,height=5.0truecm}}}
\caption{\sl
        A possible stationary state of the model 
 	with reflective boundary conditions.
	(After \cite{CBCV96}) }
\end{figure}

\def\secq{ Ref. \cite{CBCV96} }

In such simulations rotation of the particles develop (see
Fig.~\ref{spp_mosogep}) in the high density, low noise regime.  The direction
of the rotation is selected  by spontaneous symmetry breaking, thus both
clockwise and anti-clockwise spinning ``vortices'' can emerge. This rotating
state should be distinguished from the Kosterlitz-Thouless vortices
\cite{KT73}, since in our case a {\em single} vortex develops irrespective to
the system size.  In \secq we give examples of such vortices developing in
nature.

	\paragraph*{One and three dimensional SPP systems.}

Since in 1d the particles cannot get around each other, some of the
important features of the dynamics present in higher dimensions are
lost.  On the other hand, motion in 1d implies new interesting aspects
(groups of the particles have to be able to change their direction for
the opposite in an organized manner) and the algorithms used for higher
dimensions should be modified to take into account the specific crowding
effects typical for 1d (the particles can slow down before changing
direction and dense regions may be built up of momentarily oppositely
moving particles).

In a way the system studied below can be considered as a model of 
people, moving in a narrow channel. Imagine that a fire
alarm goes on, the tunnel is dark, smoky, everyone is extremely excited.
People are both trying to follow the others (to escape together) and
behave in an erratic manner (due to smoke and excitement).

Thus, in \cite{CBV99} $N$ off-lattice particles along a line of length $L$ have
been considered. The particles are characterized by their coordinate $x_i$ and
dimensionless velocity $u_i$ updated as
\begin{eqnarray}
x_i(t+\Delta t) = x_i(t) + v_0 u_i(t)\Delta t, \q
u_i(t+\Delta t) = G\Bigl(\langle u(t) \rangle_{S(i)}\Bigr) + \xi_i.
\label{EOMD}
\end{eqnarray}
The local average velocity $\langle u \rangle_{S(i)}$ for the $i$th particle is
calculated over the particles located in the interval
$[x_i-\Delta,x_i+\Delta]$, where $\Delta=1/2$.  The function $G$ incorporates
both the ``propulsion'' and ``friction'' forces which set the velocity to a
prescribed value $v_0$ on the average: $G(u)>u$ for $u<1$ and $G(u)<u$ for
$u>1$.  In the numerical simulations \cite{CBV99} one of the simplest choices
for $G$ was implemented as
\begin{equation}
G(u)=\cases{
        (u+1)/2 & for $ u > 0$ \cr
        (u-1)/2 & for $ u < 0$, \cr
     }
\end{equation}
and random initial and periodic boundary conditions were applied.

Again, the emergence of the
ordered phase was observed through a second order phase transition \cite{CBV99}
 with $\beta_{1d}=0.60\pm0.05$,
which is different from both
the the mean-field value $1/2$ and $\beta_{2d}\approx0.4$ found in 2d.
The critical line on the $\ro-\eta$ phase diagram also follows
(\ref{kappa}) with $\kappa_{1d}\approx1/4$.

In two dimensions an effective  long range interaction can build up because the
migrating particles have a considerably higher chance to get close to each
other and interact than in three dimensions (where, as is well known, random
trajectories do not overlap). The less interaction acts against ordering.  On
the other hand, in three dimensions even regular ferromagnets order.  Thus, it
is interesting to see how these two competing features change the behavior of
3d SPP systems.  The convenient generalization of (\ref{SPP2d_EOM}) for the 3d case can
be the following \cite{CVV99}:
\begin{equation}
\vec{v}_i(t+\Delta t)=v_0\hbox{ \bf N}(\hbox{ \bf N}(\langle
\vec{v}(t)\rangle_{S(i)}) + \vec\xi),
\label{EOM3D}
\end{equation}
where $\hbox{ \bf N}(\vec{u})=\vec{u}/\vert\vec{u}\vert$ and the noise
$\vec\xi$ is uniformly distributed in a sphere of radius $\eta$.

\def\reffigq{.}

Generally, the behavior of the system was found \cite{CVV99} to be similar to
that of described above.  The long-range ordered phase was
present for any $\varrho$, but for a fixed value of $\eta$, $\phi$ vanished
with decreasing $\varrho$.  To compare this behavior to the corresponding
diluted ferromagnet, $\phi(\eta,\varrho)$ was determined for $v_0=0$, when the
model reduces to an equilibrium system of randomly distributed "spins" with a
ferromagnetic-like interaction\reffigq  Again, a major difference
was found between the SPP and the equilibrium models: in
the static case the system {\it does not order} for densities below a critical
value close to 1 which corresponds to the percolation threshold of randomly
distributed spheres in 3d.

\section{Continuum approaches}\label{sec:3}
	\paragraph*{The 1d case}

Now let us focus on the continuum approaches describing the systems in terms of
coarse grained velocity and density fields. As the simplest example, let us
first investigate the 1d SPP system described in the previous section. In
\cite{CBV99} and \cite{linstab} the following equations were obtained by the
integration of the master equation of the microscopic dynamics:
\be
\partial_t\rho = -v_0 \partial_x(\rho U) + D\partial_x^2\rho
\label{CEOM1}
\ee
and
\be
\partial_tU=f(U)+\mu^2\partial_x^2 U +
\alpha{(\partial_x U)(\partial_x \rho)\over\rho} + \zeta,
\label{CEOM2}
\ee
with $\rho$ and $U$ being the coarse-grained density and dimensionless velocity 
fileds, respectively. The coefficients $v_0$, $D$, $\mu$ and $\alpha$ are 
determined by the parameters of the microscopical rules of interaction.
The function $f(U)$ is antisymmetric
with $f(U)>0$ for $0<U<1$ and $f(U)<0$ for $U>1$.
The noise $\zeta$ has a zero mean and its standard deviation is proportional
to $\rho^{-1/2}$.
The coupling term in (\ref{CEOM2}) with coefficient $\alpha$ is derived from
the Tailor expansion of the local average velocity as
\be
	\langle U \rangle (x) 
	= 
	{\int_{x-\Delta}^{x+\Delta} U(x')\rho(x')dx' 
		\over 
	\int_{x-\Delta}^{x+\Delta} \rho(x')dx'}
	=
	U + {\Delta^2\over6}\left[ \partial_x^2 U +  
2{(\partial_x U)(\partial_x \rho)\over\rho}\right] + ...
\label{1d_expand}
\ee

The nonlinear coupling term $(\partial_x U)(\partial_x \rho)/\rho$
is {\em specific} for such SPP systems: it is responsible for
the slowing down (and eventually 
the "turning back") of the particles under the influence of a larger
number of particles moving oppositely.  When two groups of particles
move in the opposite direction, the density locally increases and the
velocity decreases at the point they meet.  Let us consider a particular
case, when particles move from left to right and the velocity is locally
decreasing while the density is increasing as $x$ increases (particles
are moving towards a ``wall'' formed between two oppositely moving
groups).  The term $(\partial_x u)$ is less, the term $(\partial_x
\rho)$ is larger than zero in this case.  Together they have a negative
sign resulting in the slowing down of the local velocity.  This is a
consequence of the fact that there are more slower particles (in a given
neighborhood) in the forward direction than faster particles coming from
behind, so the average action experienced by a particle in the point $x$ 
slows it down.

For $\alpha=0$ the dynamics of the velocity field $U$ is independent of $\rho$,
and Eq.~(\ref{CEOM2}) is equivalent to the time dependent Ginsburg-Landau
$\Phi^4$ model of spin chains, where domains of opposite magnetization develop
at finite temperatures.  As numerical simulations demonstrate
\cite{CBV99,linstab}, for large enough $\alpha$ and low noise the initial
domain structure breaks and the system is organized into a single major group
traveling in a spontaneously selected direction.  
This elimination of the domain structure, i.e., the ordering 
is strongly connected to the linear instability of the
domains \cite{CBV99,linstab} for large enough $\alpha$.

	\paragraph*{Hydrodynamics in two and higher dimensions.}

To get an insight into the possible analytical treatment of SPP
systems in higher dimensions, 
the Navier-Stokes equations can be applied
for a fluid of SPPs, as described in \cite{CsCz97}.
The two basic equations governing the dynamics of the {\em noiseless}
``self-propelled fluid''
are the continuity equation
\begin{equation}\label{cont0}
\partial_t \rho + \nabla (\rho {\vec v})= 0 
\end{equation}
and the equation of motion
\begin{eqnarray}\label{eom0} 
\partial_t {\vec v} + ({\vec v}\cdot \nabla) {\vec v} &=&
{\vec F}({\vec v}) - {1\over\tau_1} {\vec v} 
+{1\over\tau_2}\left(\langle {\vec v} \rangle - v\right)
-{1\over \rho^*} \nabla p + \nu \nabla^2 {\vec v},
\end{eqnarray}
where $\rho^*$ is the effective density of the particles,
$p$ is the pressure, $\nu$ is the kinematic viscosity,
${\vec F}({\vec v})$ is the {\it intrinsic} driving force of biological
origin and
$\tau_1$, $\tau_2$ are time scales associated with velocity relaxation
resulting from interaction with the environment and the surrounding SPPs, respectively.
Let us now go through the terms of Eq.\ \ref{eom0}.

The self-propulsion can be taken
into account as a constant magnitude force acting
in the direction of the velocity of the particles as
\begin{equation}
{\vec F} = {v_0\over\tau_1}{{{\vec v}}\over{\vert {{\vec v}}
\vert}},
\label{sdrive}
\end{equation}
where $v_0$ is the speed determined by the balance of the propulsion
and friction forces, i.e., $v_0$ would be the speed of a homogeneous
fluid.  

Taking Taylor expansion of the expression 
${1\over\tau_2}\left(\langle {\vec v} \rangle - v\right)$
describing local velocity alignment
yields
\begin{eqnarray}
\langle {{\vec v}} \rangle_{\epsilon} - {\vec v} &=&
{
\int_{\vert {\vec \xi} \vert < \epsilon} \hbox{d} {\vec \xi} \Bigl(
{{\vec v}}\rho+
({\vec \xi}\nabla) {{\vec v}} \rho +{1\over 2}({\vec \xi}\nabla)^2
{{\vec v}}\rho
+ \dots \Bigr)
\over
\int_{\vert {\vec \xi} \vert < \epsilon} \hbox{d} {\vec \xi} \Bigl(
\rho + ({\vec \xi}\nabla)\rho +{1\over 2}({\vec
\xi}\nabla)^2\rho
+ \dots \Bigr) } - {\vec v} \nonumber\cr 
& =&  {\epsilon^2\over 6}\Bigl(
{\nabla^2 ({{\vec v}}\rho)\over \rho} -
{{\vec v}}{\nabla^2 \rho \over \rho} \Bigr) + \dots \nonumber\q
&=&
{\epsilon^2\over 6}\Bigl( \nabla^2{{\vec v}} +
2(\nabla {{\vec v}}) {\nabla\rho\over\rho} \Bigr) + \dots
\label{DIFF1}
\end{eqnarray}
similarly to the 1d case (\ref{1d_expand}). In the following we will consider cases where the 
density fluctuations are forced to be small. Hence the velocity-density
coupling term in  (\ref{DIFF1}) is negligible, which is a major difference 
compared to the previously studied 1d system. Thus,
if the relative density changes are small, the velocity alignment can be
incorporated with an effective viscosity coefficient $\nu^*$
into the viscous term of (\ref{eom0}), 

In the simplest cases the pressure can be composed of an effective 
``hydrostatic'' pressure, and an externally applied pressure as
\begin{equation}\label{press}
p = g^*\rho + p_{\rm ext}
\end{equation}
where $g^*$ is a parameter related to the compressibility of the fluid.

Combining  (\ref{cont0}), (\ref{eom0}), (\ref{sdrive}) and
(\ref{press})  one gets the following final form for the equations
of the SPP flow:
\begin{equation}
\label{cont}
\partial_t \rho + \nabla (\rho {\vec v})= 0, 		
\end{equation}
and
\be
\label{eom}
\partial_t {\vec v} + ({\vec v}\cdot \nabla) {\vec v} =
{v_0\over\tau_1}{{{\vec v}}\over{\vert {{\vec v}} \vert}} 
- {1\over\tau_1} {\vec v}
+ \nu^* \nabla^2 {\vec v}
-{g^*} \nabla \rho +
-{1\over\rho^*}\nabla p_{\rm ext}.
\ee

It is useful to introduce the characteristic length
$\lambda=\sqrt{\nu^*\tau_1}$
and characteristic time 
$T={\lambda\over v_0}$
and write (\ref{cont}) and  (\ref{eom}) in a dimensionless form of
\begin{equation}
\label{cont1a}
\partial_{t'} \rho + \nabla' (\rho {\vec v}')= 0, 	
\end{equation}
\be
\label{eom1a}
{\tau_1\over T}
	\left(\partial_{t'} {\vec v}' + ({\vec v}'\cdot \nabla') {\vec v}' 
	\right)
	= {{\vec v}'\over{\vert {\vec v}' \vert}} - \vec{v}' + \nabla'^2\vec{v}'
	- g'_1 \nabla' \rho + g'_2 \nabla' p_{\rm ext}.
\ee
with $t'=t/T$, $x'=x/\lambda$, $v'=v/v_0$ and the $\nabla'$ operator derivates
with respect to ${\vec r}'={\vec r}/\lambda$. We also introduced the notations
$g'_1=T\tau_1g^*/\lambda^2$ and $g'_2=g'_1/g^*$.
In the following we will drop the prime for simplicity.

For certain simple geometries 
it is possible to obtain analytical stationary solutions \cite{CsCz97}
of the above equations supposing incompressibility ($\rho=const.$) and
see that the minimal size 
of a vortex is in the order of $\lambda$, i.e., one in dimensionless units.
Therefore, if the dimensionless system size, $R\gg 1$ then initially
many Kosterlitz-Thouless-like vortices are likely to be present in the system.

\begin{figure}
\centerline{\psfig{figure=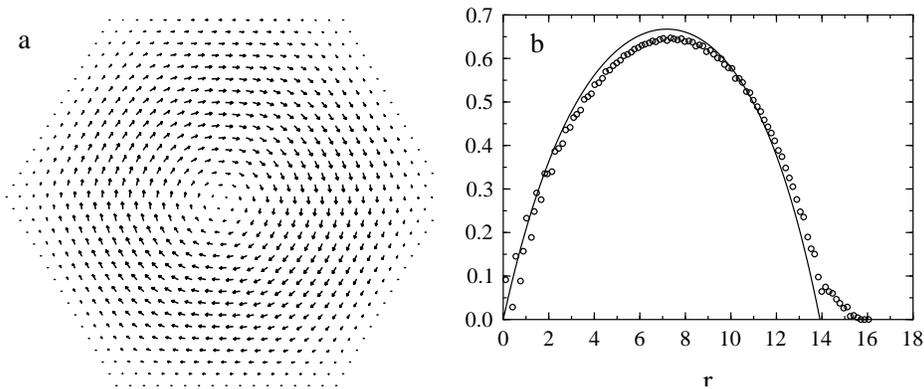,width=14cm,angle=-90}}
\caption{\sl
	(a) A numerically generated vortex for dimensionless system size
	$R=4.4$ and $g_1=750$.  
	The length of the arrows is proportional to the local velocity,
	while their thickness is proportional to the density.
	(b) The measured (circles) and the theoretical (solid line) velocity
	profile for the vortex shown in (a).
	(After \cite{CsCz97}.)
}
\label{numvort}
\end{figure}

\begin{figure}
\centerline{\psfig{figure=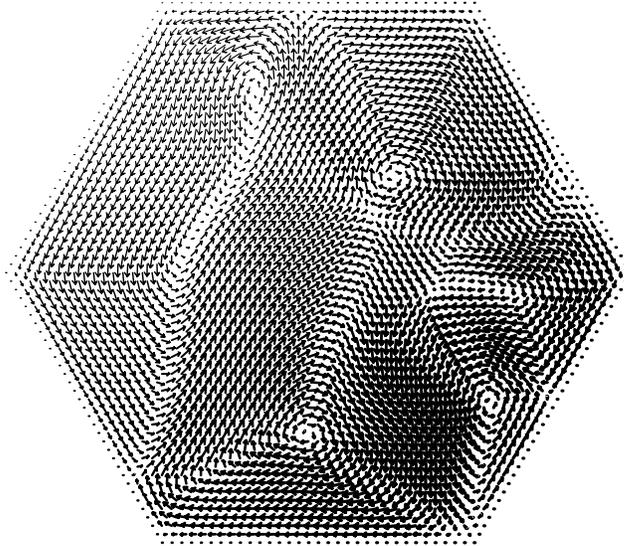,width=8cm}}
\caption{\sl 
	Transient multiple vortex state at $R= 25.3 , g_1= 128$. 
	(After \cite{CsCz97}.)}
\label{xxvortex}
\end{figure}

Numerical solutions of 
(\ref{cont}) and (\ref{eom}) were also calculated in \cite{CsCz97} with 
finite compressibility and $p_{\rm ext}=0$.
The only remaining dimensionless quantity characterizing the system
is $\tau_1/T$ which relates the various relaxation times and $v_0$. The system
can  be considered overdamped if $\tau_1\ll T$ holds.
Fig.\ \ref{numvort} shows the stationary state for the equations
in the high compressibility ($g_1=750$) and overdamped limit.
The length and direction of the arrows show the velocity, while
the thickness is proportional to the local density of the fluid.
In Fig.\ \ref{numvort}b the radial velocity distribution is presented for
the vortex
shown in Fig.\ \ref{numvort}a together with the velocity profile obtained
analytically in \cite{CsCz97}. 
Rather good agreement is seen; the differences are
due to the
fact that the numerical system is not perfectly circular.
In simulations with $R\gg1$ multi-vortex states develop
(Fig.\ \ref{xxvortex}), which transform into a single vortex after a
long enough time.

Thus, we demonstrated that -- in contrast to the 1d case -- the ordering in 
2d is not due to the density-velocity coupling term in the expansion of $\langle
v \rangle$: the viscosity and the internal driving force terms are sufficient
to destabilize the originally present vortices and organize
the system into a globally ordered phase. The stability of that ordered phase
against a finite amount of fluctuations has been shown by Tu and Toner in
\cite{TT95,TT98} and discussed in this volume, too.

\end{document}